\documentclass{aa}  

\usepackage{balance}
\usepackage{graphicx}
\usepackage{txfonts}
\usepackage{amsmath}
\usepackage{booktabs,caption}
\usepackage[flushleft]{threeparttable}

\def\msun{M$_\odot$}

\def\rg{$R_{\rm g}$}

\def\mdot{$\dot{m}_{\rm Edd}$}
\def\mbh{$M_{\rm BH}$}
\def\m8p5{M$_{8.5}$}
\def\nttemp{$T_{\rm NT}(R)$}

\def\risco{$R_{\rm ISCO}$}
\def\rtr{$R_{\rm tr}$}
\def\rtroverisco{$R_{\rm tr}/R_{\rm ISCO}$}
\def\spin{$a^\ast$}
\def\kynsed{{\tt KYNSED}}

\def\lumobs{$L_{\rm 3000}$}
\def\hlrobs{$R_{\rm 1/2}$}
\def\hlrobsa{$R_{\rm 1/2,a}$}
\def\hlrobsb{$R_{\rm 1/2,b}$}

\def\lummod{$L_{\rm 3000,mod}$}
\def\hlrmod{$R_{\rm 1/2,mod}$}

\def\ahlro{$a_{\rm 1/2,obs}$}
\def\bhlro{$b_{\rm 1/2,obs}$}

\def\alumo{$a_{\rm L,obs}$}
\def\blumo{$b_{\rm L,obs}$}
\def\alumotwo{$a_{\rm L,obs-2}$}
\def\blumotwo{$b_{\rm L,obs-2}$}

\def\lxm{$L_{\rm Xmod,10-50}$}
\def\lxobs{$L_{\rm Xobs,10-50}$}

\usepackage[breaklinks, colorlinks, citecolor=blue, urlcolor = blue]{hyperref}

\makeatletter
\renewcommand*\aa@pageof{, page \thepage{} of \pageref*{LastPage}}
\makeatother

\begin{document} 

 \title{X-ray illuminated accretion discs and quasar microlensing disc sizes \thanks{Tables 4 and 5 are only available in electronic form
at the CDS via anonymous ftp to cdsarc.u-strasbg.fr (130.79.128.5)
or via http://cdsweb.u-strasbg.fr/cgi-bin/qcat?J/A+A/}}

\author{I. E. Papadakis \inst{\ref{inst1},\ref{inst2}}
\and
M. Dov{\v c}iak \inst{\ref{inst3}}
\and 
E. S. Kammoun \inst{\ref{inst4},\ref{inst5}}
          }
\institute{
Department of Physics and Institute of Theoretical and Computational Physics, University of Crete, 71003 Heraklion, Greece
\email{\href{mailto:jhep@physics.uoc.gr}{jhep@physics.uoc.gr}} \label{inst1} 
\and 
Institute of Astrophysics, FORTH, GR-71110 Heraklion, Greece\label{inst2}
\and
Astronomical Institute of the Czech Academy of Sciences, Bo{\v c}n{\'i} II 1401, CZ-14100 Prague, Czech Republic \label{inst3} 
\and 
IRAP, Universit\'e de Toulouse, CNRS, UPS, CNES 9, Avenue du Colonel Roche, BP 44346, F-31028, Toulouse Cedex 4, France\label{inst4}
\and
INAF -- Osservatorio Astrofisico di Arcetri, Largo Enrico Fermi 5, I-50125 Firenze, Italy\label{inst5}
}   

   \date{Received ; accepted }

 
  \abstract
   {}
   {We study the half-light radius versus black hole mass as well as the luminosity versus black hole mass relations in active galactic nuclei (AGN) when the disc is illuminated by the X-ray corona.}
   { We used \kynsed, a recently developed spectral model for studying broadband spectral energy distribution in AGN. We considered non-illuminated Novikov-Thorne discs and X-ray illuminated discs based on a Novikov-Thorne temperature radial profile. We also considered the case where the temperature profile is modified by a colour-correction factor. In the case of X-ray illumination, we assumed that the X-ray luminosity is equal to the accretion power that is dissipated to the disc below a transition radius and we computed the half-light radius and the disc luminosity for many black hole masses, as well as a wide range of accretion rates, black hole spins, X-ray luminosities and heights of the corona.}
   {The half-light radius of X-ray illuminated radii can be up to $\sim 3.5$  times greater than the radius of a standard disc, even for a non-spinning black hole, based on a wide range of model parameters -- as long as the transition radius is larger than six times the radius of the innermost stable circular orbit and the coronal height is greater than $\sim 40$ \rg. These requirements are due to the fact that the absorbed X-rays act as a secondary source of energy, increasing the disc temperature, mainly at large radii. Non-illuminated discs are consistent with observations, but only at the 2.5$\sigma$ level. On the other hand, X-ray illuminated discs can explain both the half-light radius-black hole mass as well as the luminosity-black hole mass relation in AGN, for a wide range of physical parameters. The range of the parameter space is broader in the case where we consider the colour-correction factor. X-ray illuminated discs can explain the data when we observe  gravitationally lensed quasars mainly face-on, but also if the mean inclination angle is 60\degr. In addition, we show that the observed X-ray luminosity of the gravitationally lensed quasars is fully consistent with the X-ray luminosity that is necessary for heating the disc.}
   {X-ray disc illumination was proposed many years ago to explain various features that are commonly observed in the X-ray spectra of AGN. Recently, we showed that X-ray illumination of accretion disc can also explain the observed UV/optical time-lags in AGN, while in this work, we show that the same model can also account for the quasar micro-lensing disc size problem. These results support the hypothesis of the disc X-ray illumination in AGN.}


   \maketitle %

\section{Introduction}

It is widely accepted that active galactic nuclei (AGN) are powered by the accretion of matter into a supermassive black hole (BH). \citet{Shakura73} and \citet[][(hereafter NT)]{Novikov73} are standard accretion disc models that are frequently used to explain the accretion process in these objects. 
Since AGN accretion discs cannot be spatially resolved with ordinary telescopes, we have to rely on techniques other than imaging to measure their size.
An example of this is gravitational microlensing due to stellar-mass objects, such as stars in the lensing galaxy, which results in flux fluctuations. Their amplitude depends on the size of the emitting source. Thus, the results from optical microlensing observations can be used to measure the accretion disc size. Using this technique, disc sizes appear to be systematically larger than the standard model predictions by a factor of $\sim 2-4$ \citep[e.g.][]{Pooley07,Dai10,Morgan10,JimenezVicente12,Blackburne14,Munoz16,Motta17}. Various physical explanations have been proposed over the past few years to explain this difference  \citep[e.g.][]{Dexter11, Abolmasov12, Jiang16, Hall18,li19}. 

AGN are strong X-ray emitters and X-ray illumination of the accretion disc was proposed many years ago in order to explain the Fe K$\alpha$ emission line around 6.4~keV and the spectral hardening above $\sim 10$ keV \citep[e.g.][]{Pounds90,George1991,Nandra91}. Since then, X-ray illumination models of the inner disc have developed considerably and can account for various features in the X-ray spectra of AGN (i.e. the iron line shape, the Compton hump, and the soft excess), as well as the detected time-lags of the soft band variations with respect to the hard photons at high frequencies \cite[e.g.][]{Fabian2009}. 

\begin{table*}
\centering
\caption[]{BH mass, half-light radius, $I-$band magnitudes, redshift, luminosity distance, and luminosity for the sources in the sample.}
\begin{tabular}{lcccccl}
\hline \hline
 Name & log(\mbh) & log(R$_{\rm 1/2}$) & $I_{\rm corr}$ & $z$ & $D_{\rm L}$   &  log(L$_{3000}$) \\
     &    (M$_{\odot})$ & (cm)  &   (mag)             &     & (Gpc)  &  (ergs/s)  \\
\hline
QJ~0158      &    8.2~$^a$         & $15.14\pm0.3$     &  19.09$\pm 0.12$ & 1.290 & 9.3  & 45.32$\pm0.05$\\
HE~0435      &    8.7~$^a$         & $15.94\pm0.6$     &  20.76$\pm 0.25$ & 1.689 & 13   & 44.94(45.10)$\pm0.10$ \\
SBS~0909     &    8.5~$^b$         & $15.51\pm0.30~^e$ &  $-$             & 1.378 & --   &  -- \\
SDSS~0924    &    8.0~$^a$         & $15.24\pm0.35$    &  21.24$\pm 0.25$ & 1.523 & 11.4 & 44.64$\pm0.10$ \\
FBQ~0951     &    8.95~$^a$        & $16.34\pm0.35$    &  17.16$\pm 0.11$ & 1.246 & 8.9  & 46.00$\pm0.04$ \\
Q~0957       &    9.0~$^b$         & $16.42\pm0.50~^e$ &  $-$             & 1.416 & --   &  --    \\
SDSS~1004    &    8.6~$^a$         & $15.14\pm0.3$     &  20.97$\pm 0.44$ & 1.738 & 13.4 & 44.89(45.0)$\pm0.18$ \\
HE~1104      &    9.4~$^b$         & $16.14\pm0.25$    &  18.17$\pm 0.31$ & 2.319 & 19.1 & 46.31$\pm0.12$ \\
PG~1115      &    9.1~$^b$         & $16.84\pm0.35$    &  19.52$\pm 0.27$ & 1.733 & 13.4 & 45.47$\pm0.11$  \\
RXJ~1131     &    7.8~$^a$         & $15.54\pm0.2$     &  20.73$\pm 0.11$ & 0.654 & 4.0  & 43.94$\pm0.04$  \\
SDSS~1138    &    7.7~$^b$         & $15.14\pm0.6$     &  21.97$\pm 0.19$ & 2.445 & 20.4 & 44.85$\pm0.08$  \\
SBS~1520     &    8.95~$^a$        & $15.94\pm0.2$     &  18.92$\pm 0.13$ & 1.855 & 14.5 & 45.78$\pm0.05$  \\
WFI~2033     &    8.6~$^c$         & $16.26\pm0.30~^e$ &   $-$            & 1.661 & --   &  --  \\
WFI~2026     &    9.4~$^d$         & $15.98\pm0.31~^e$ &   $-$            & 2.223 & --   & -- \\
Q~2237       &    8.7.~$^b$        & $15.84\pm0.3$     &  17.90$\pm 0.44$ & 1.695 & 13   & 46.09(46.3)$\pm0.18$  \\ 
\hline
\end{tabular}
\tablefoot{BH mass measurements are taken from \cite{Morgan10} ($^a$), and \cite{Assef11} ($^b$). When not available, we use the measurements of \cite{Sluse12} ($^c$), and \cite{Cornachioneb} $(^d)$. The logarithm of $R_{\rm 1/2}$ (third column) are taken from \cite{Morgan10}, except from ($^e$) which are taken from \cite{Cornachionea}. Fourth column lists the (unmagnified) $I-$band magnitudes taken from \cite{Morgan10}. Redshift and luminosity distances (fifth and sixth column, respectively) are taken from NED\footnote{The NASA/IPAC Extragalactic Database (NED) is funded by the National Aeronautics and Space Administration and operated by the California Institute of Technology}, assuming H$_0 = 67.8$\,km/sec/Mpc, $\Omega_{\rm matter} = 0.308$, $\Omega_{\rm vacuum} = 0.692$.}
\label{table:data}
\end{table*}

The X-ray illumination of the accretion disc makes one implicit prediction. Since most of the X-rays will be absorbed by the disc, they will add to its heating. This externally supplied energy will appear in the form of thermal emission at UV/optical wavelengths, which should be variable (since X-rays are highly variable in AGN) and correlated with the X-rays, with some delay. \citet[][(hereafter K21)]{Kammoun21a} studied in detail the response of a NT disc to X-ray illumination by a point-like source, located above the central BH, taking into account all relativistic effects, and using the latest X-ray reflection models. The observed UV/optical time-lags, which are measured from recent multi-wavelength monitoring campaigns of AGN, appear to imply a disc radius that is a few times times larger than the prediction from standard thin-disc theory. However,  \cite{Kammoun19lags,Kammoun21b} showed that when treated properly, the time-lags within the UV/optical band, in the case of X-ray illuminated discs, are in agreement with the observations. The observed time-lags in a few energy bands are larger than expected, most probably due to the significant contribution, and longer response, from the broad line region continuum emission in these bands \citep[see e.g.][]{Korista01, Korista19, Guo2022, Netzer2022}. We note that the K21 model can also explain the observed power-spectra in the optical/UV bands of NGC~5548 \citep{Panagiotou20}.

Recently, \citet[][(hereafter D22)]{Dovciak21} published a new spectral model, named \kynsed\footnote{\url{https://projects.asu.cas.cz/dovciak/kynsed}},  which calculates the broadband and the optical/UV/X-ray spectral energy distribution (SED) in AGN when the accretion disc is illuminated by an X-ray corona. The model assumes that the accretion power dissipated in the disc below a transition radius, $r_{\rm tr}=R_{\rm tr}/$\rg\footnote{\rg$=G$\mbh$/c^2$ is the gravitational radius of a BH with a mass of \mbh. Distances in units of \rg\ are written in the lower case letter $r$.}, is transferred to the X-ray source (by an as yet unspecified physical mechanism). In this way, the model sets a direct link between the accretion disc and the X-ray source. D22 showed that this model can fit the mean SED of NGC~5548 well, using data from the STORM multi-wavelength monitoring campaign from 2014 \citep{fausnaugh16}.

In this work, we investigate whether X-ray illuminated discs can resolve the discrepancy that appears to exist between the size of the continuum emission region as measured by microlensing observations and the luminosity-based disc sizes. We use \kynsed\ to compute the radial disc profiles, the disc half-light radius, and the luminosity for a wide range of physical parameters and we compare the resulting half-light radius versus BH mass as well as the luminosity versus BH mass relations with the observations. 

The paper is organised as follows. In \S\ref{sec:data} we present the sources and the data we use in this paper. In \S\ref{sec:standard} we discuss the half-light radius in standard NT discs (for low and high spins, taking into account relativistic effects as well as the colour-correction terms), while in \S\ref{sec:illuminated} we discuss the disc half-light radius in NT discs that are illuminated by X-rays. In \S\ref{sec:model} we present  model half-light radius and luminosity for a wide range of physical parameter values and  in \S\ref{sec:results} we compare the model predictions with the observations. Finally, we discuss our results in \S\ref{sec:discuss}.

\section{Quasar micro-lensing variability sample}
\label{sec:data}

We consider the quasar sample of \cite{Morgan10}, which consists of 11 objects with accretion disc microlensing measurements using multi-epoch light curves. We added half-light radius measurements for four objects from \cite{Cornachionea}. The sources and data are listed in Table \ref{table:data}. Black hole mass (\mbh) measurements are taken from the literature.  They are based on the observed \ion{C}{IV} and \ion{Mg}{II} line widths (H$\beta$ for RXJ 1131-1231), and locally calibrated virial relations for black hole masses. The statistical uncertainties on the BH mass measurements are quite small, but the errors are dominated by systematic uncertainties. We assumed a systematic uncertainty of 0.3 dex  \citep[e.g.][]{Vestergaard06} for all the BH mass measurements listed in Table \ref{table:data}. We note that the systematic uncertainty may be larger and, in fact, the use of \ion{C}{IV} width may result in biased BH mass estimates \citep{Trakhtenbrot2012}. 

The half-light radius measurements (\hlrobs) are taken from \cite{Morgan10} and \cite{Cornachionea}, and are scaled to 2500~\AA\ in the quasar rest frame. The values listed in Table \ref{table:data} are not corrected for inclination effects. The disc size measurements are set by the projected area of the disc. We considered two possibilities: a) quasars in the sample are uniformly distributed over the range $\cos(\theta=45^\circ) < \cos(\theta) < \cos(\theta=0^\circ)$ and b) quasars are uniformly distributed over the range $\cos(\theta=90^\circ) < \cos(\theta) < \cos(\theta=0^\circ)$, where $\theta$ is the disc inclination angle. Under case (a) quasars are seen preferentially face-on due to the presence of the molecular torus, while the second possibility corresponds to the hypothesis that there is no torus in these systems. In case (a), $\langle \cos(\theta)\rangle=0.85$ (i.e. $\theta\approx 30^\circ$) and we have to add $\log\left[\sqrt{1/\cos(30^\circ)}\right]=0.03$ to log(\hlrobs), to account for the inclination effects, while $\langle \cos(\theta)\rangle=0.5$ ($\theta = 60^\circ$) in case (b); and we have to add $\log\left[\sqrt{1/\cos(60^\circ)}\right] = 0.15$ to the half-radius measurements.

The fourth column in Table \ref{table:data} list $I-$band magnitudes taken from \cite{Morgan10}. The sources were observed with the F814W filter on {\it HST}, which corresponds to a (median) rest-frame wavelength of $\sim 3000$~\AA\ for the sources in the sample. The magnitudes are corrected for the lens magnification, but not for the contribution of the Balmer continuum. We applied corrections for the Galactic absorption using the $E(B-V)$ measurements of \cite{schlafly11}, and assuming $R_V=3.1$. \cite{Chen12} report X-ray absorption by neutral material in HE 0435 and SDSS 1004 (N$_{\rm H}\sim 4 \times 10^{20}$ cm$^{-2}$, in both cases), and in Q 2237 ( N$_{\rm H}=7^{+2}_{-2}\times 10^{20}$ cm$^{-2}$). We used the ratio of N$_{\rm H}/E(B-V)=8.3\times 10^{21}$ \citep{Liszt21} to compute the intrinsic $E(B-V)$ in these sources, and then the extinction curve of \cite{Czerny04} to correct the emitted flux at 3000~\AA\footnote{This approach assumes absorption intrinsic to the host galaxy, although this may be due to the lensing galaxy. Nevertheless,  the difference is not significant, given the relatively small $N_{\rm H}$.}. We converted the absorption-corrected magnitudes to fluxes adopting an $I-$band magnitude zero point of 2409 Jy (C. Morgan, priv. comm). We used the luminosity distances listed in Table \ref{table:data} to compute the rest frame 3000~\AA\ luminosity (in erg/s), \lumobs, from the observed $I-$band fluxes. The luminosity measurements are listed at the last column of Table \ref{table:data}. Numbers in parentheses indicate the intrinsic luminosity when accounting for the extragalactic absorption for three sources, as explained above.

\begin{figure}
\centering
\includegraphics[width=0.9\linewidth]{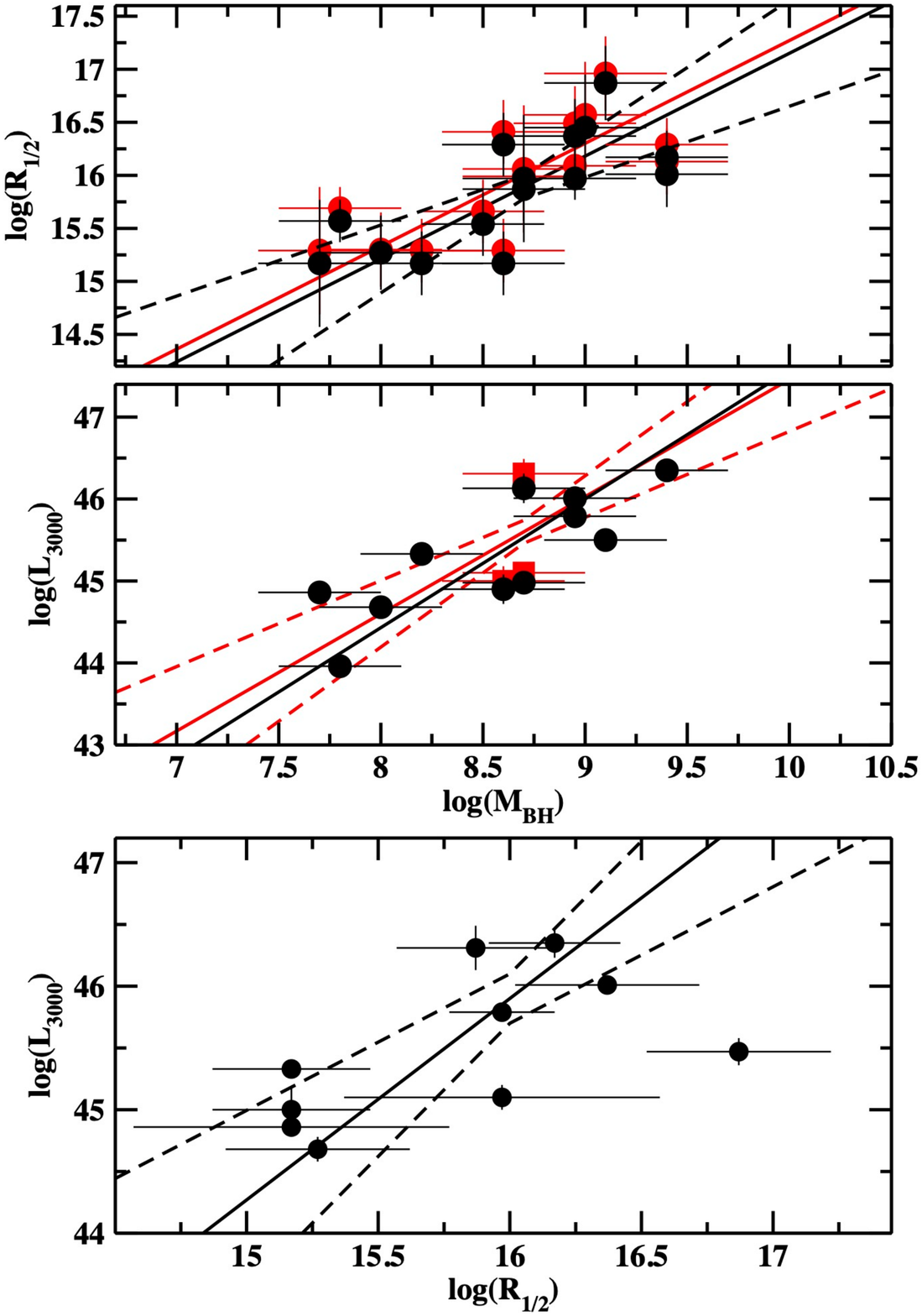}
\caption{Log(\hlrobs) and log(L$_{3000})$ vs. log(\mbh) (upper and middle panels, respectively). Black and red circles in the top panel show \hlrobs\, assuming an inclination angle of 30\degr\ and 60\degr\,, respectively. Red points in the middle plot show data after correcting for intrinsic absorption as well. A plot of log(L$_{3000}$) vs logarithm of half half-light radius is shown in the bottom panel. Solid lines show the best fit to the data, while dashed lines show the $1\sigma$ confidence regions (see text for details).}
\label{fig:data}
\end{figure}

\begin{table}
\centering
    \caption{Summary of the best-fit results to the observed relations shown in Fig.\,\ref{fig:data}, as reported in \S\ref{sec:data}.}
\begin{tabular}{llll}
\hline \hline
Data sets       &               Intercept   &  Slope   & $\chi^2$/dof   \\ \hline
\hlrobsa\, vs \mbh\ & 15.89$\pm0.11$ & 0.97$\pm0.30$ & 11.5/13  \\
\hlrobsb\, vs \mbh\,  & 16.01$\pm0.11$   & 0.97$\pm0.30$ & 11.5/13\\
\lumobs\, vs \mbh                          &  45.53$\pm0.16$ & 1.57$\pm 0.34$ & 11.6/9     \\
\lumobs\, vs \mbh$^{\dagger}$                         &  45.60$\pm0.14$ & 1.43$\pm 0.38$ & 11.8/8   \\
\lumobs\, vs \hlrobsa\,                       &  45.90$\pm0.20$ & 1.60$\pm0.55$ &16.5/8    \\ \hline
\end{tabular}
\tablefoot{$^\dagger$Best-fit results to the luminosity vs BH mass relation when we exclude RXJ 1131 and we consider intrinsic absorption to three quasars (i.e. results listed in this row are the \alumotwo\, and \blumotwo\, values reported in the text).}
\label{table:bestfit}
\end{table}

Figure \ref{fig:data} shows the logarithm of the observed half-light radius, log(\hlrobs), and luminosity, log(\lumobs), plotted vs log(\mbh) (top and middle panel, respectively). Black and red circles in the top panel show the half-light measurements when we assume case (a) and case (b) uniform distribution of disc inclination angles (\hlrobsa\, and \hlrobsb, respectively). The red squares in the middle panel of Fig.\,\ref{fig:data} show the three luminosity measurements which are corrected for extragalactic reddening. The bottom panel in the same figure shows a plot of $\log(L_{3000})$ vs $\log($\hlrobsa), i.e. the half-light radius when we assume a median inclination of 30 degrees.

There is a clear positive correlation between disc size and luminosity with BH mass. We fit the data in the upper two panels of Fig.\,\ref{fig:data} with a straight line taking the form: $Y=a+b[X-8.7]$, where $X$=log(\mbh), and $Y$=log(\hlrobs). We normalized \mbh\ at $=5\times 10^8\,\rm M_{\odot}$ (i.e.  log(\mbh)=8.7) to minimize the error on $a$ and $b$. The fit was done in the log--log space, using the {\tt fiteyx} routine of \cite{Press2007}, which takes into account the error on both variables. The best-fit results in case (a) are: \ahlro=15.89$\pm0.11$, \bhlro=0.97$\pm0.30$, and $\chi^2=11.5/13$ degrees of freedom (dof). The best-fit normalization in case (b) is 16.01$\pm0.11$, while the best-fit slope and the best-fit $\chi^2$ remain the same as in case (a). The best-fit results 
are also listed in Table\,\ref{table:bestfit}. 

The black and red solid lines in  the upper panel of Fig.\,\ref{fig:data} show the best-fit linear model to the log(\hlrobsa)/log(\hlrobsb) vs log(\mbh) data. The dashed black lines show the $1\sigma$ confidence limits of the best-fit relation to the log(\hlrobsa) vs log(\mbh) data, taking into account both the error on $a$ and $b.  $ We note that we do not show the respective lines for the log(\hlrobsb) versus log(\mbh) data to maintain clarity.

\begin{figure*}
\centering
\includegraphics[width=0.49\linewidth]{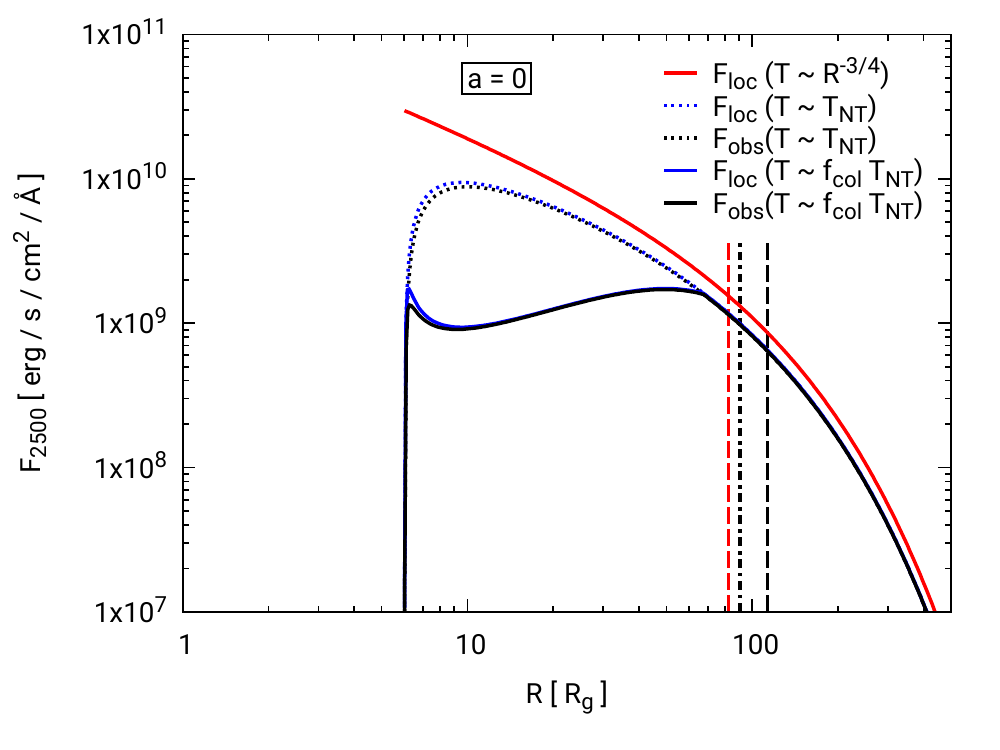}
\includegraphics[width=0.49\linewidth]{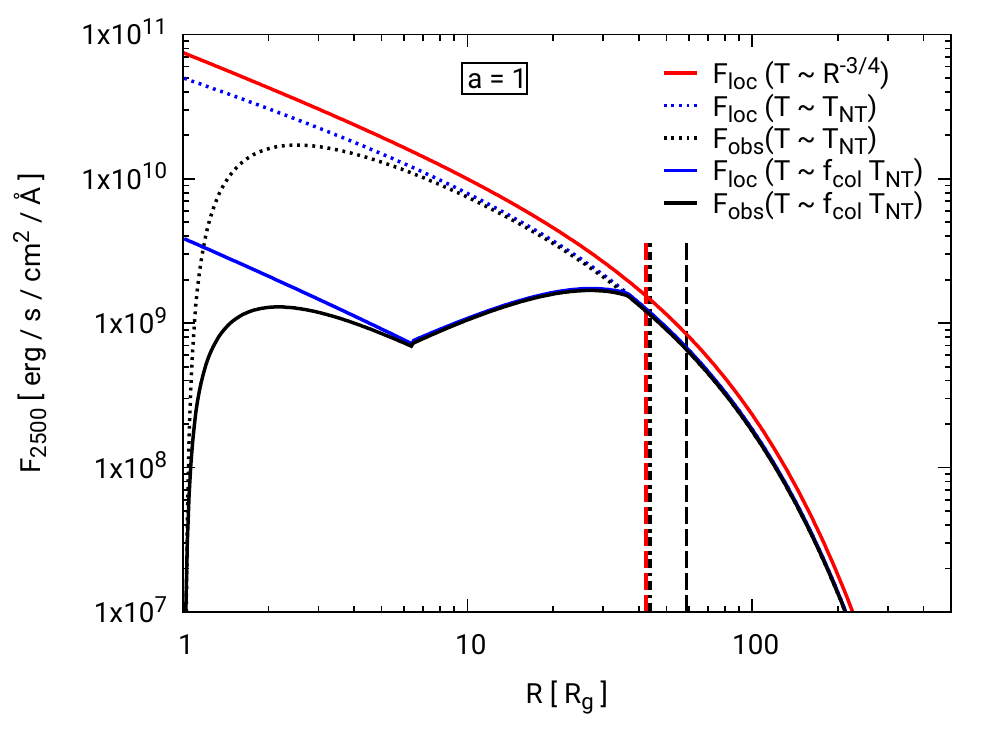}
\caption{Radial disc flux profile (from one side of the disc) at 2500~\AA\ for \mbh$=10^8$\msun, \mdot$=0.2$, $\theta=40^{\circ}$, \spin$=0$ and 1 (left and right panels, respectively).
We consider different radial profiles: simple approximations and the proper NT radial temperature dependence (red and dotted blue lines); the radial profile when we include the temperature colour correction factor of \cite{Done2012} (solid blue line); and when all relativistic effects are included (black lines). The latter are the profiles observed at infinity. The vertical lines show the location of \hlrmod for the various radial profiles (see \S\ \ref{sec:standard} for details).}
\label{fig:radprofl}
\end{figure*}

\begin{figure}
\centering
\includegraphics[width=1\linewidth]{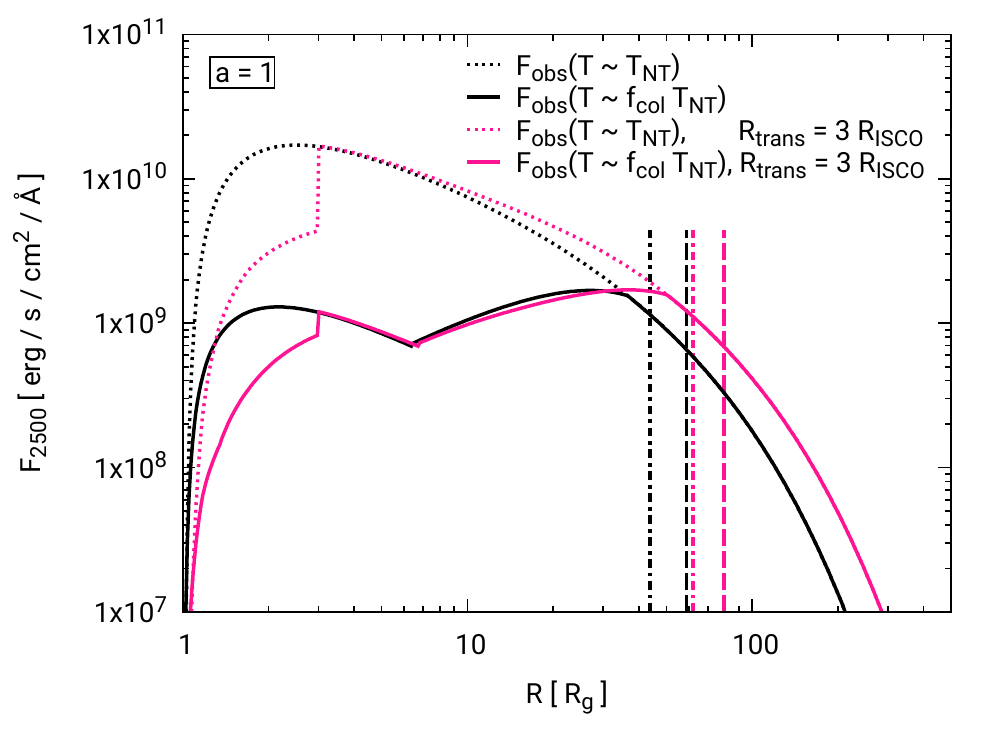}
\caption{ Magenta lines show the radial flux profile (for one side of the disc) at 2500~\AA\ (as seen at infinity) when the disc is illuminated by an X-ray corona at $h=20$~\rg\, and \rtr=3\risco\ for BH with \spin=1. All other model parameters are the same as in Fig.\,\ref{fig:radprofl}. Dotted and solid lines indicate the profile for a NT disc and when we assume the colour correction factor of \cite{Done2012}, respectively.
The black lines are the same as the black lines in Fig.~\ref{fig:radprofl}, and are shown for comparison. The vertical dashed lines show the location of \hlrmod\ for each radial profile.}
\label{fig:radprof-rtrans}
\end{figure}

The best-fit results for the log(\lumobs) versus log(\mbh) plot are: \alumo=$45.53\pm0.16$, \blumo$=1.57\pm0.34$ ($\chi^2=11.6/9$ dof). The black solid line in the middle panel of Fig.\,\ref{fig:data} shows the best-fit line. The red solid line in the same panel shows the best-fit line when we consider the red points and we exclude the RXJ 1131 measurement. This is the source with the lowest redshift among all sources in the sample, and its observed flux corresponds to rest frame flux at $\sim 5000$~\AA, which is quite longer than the rest frame wavelength of the other sources. The best-fit results in this case are \alumotwo=$45.60\pm0.14$, \blumotwo$=1.43\pm0.38$ ($\chi^2=11.1/8$ dof) and are consistent, within 1$\sigma$, with the results from the best-fit to the original data. The dashed red lines show the $1\sigma$ confidence limits of the best-fit relation in the latter case, taking into account the error on \alumotwo\, and \blumotwo.

The bottom panel in Fig.\,\ref{fig:data} shows that \lumobs\, and \hlrobs\, are also positively correlated. We fit the data with a straight line of the form: log(\lumobs)$=a+b[\log($\hlrobsa$)-16]$. We normalized the \hlrobsa\, data to \hlrobsa=10$^{16}$ cm  to minimize the error on $a$ and $b$. The best-fit results are: $a_{\rm L,R_{1/2}}=45.9\pm 0.2$, $b_{\rm L,R_{1/2}}=1.6\pm 0.55$, and $\chi^2=16.5$ for 8 dof. The solid and dashed lines show the best-fit, and the 1$\sigma$ confidence region. The best-fit results indicate that the main-driver for the \lumobs vs  \hlrobs\, relation is the positive relation of these parameters with BH mass. If \hlrobs$\propto$\mbh\, and \lumobs$\propto$(\mbh$)^{1.6}$, then we expect \lumobs$\propto$(\hlrobs)$^{1.6}$, as observed. Given this, and the fact that the error of the best-fit parameters to the \lumobs\, vs \mbh\, relation is larger than in the other two relations (mainly the error on the slope), we concentrate below on the study of the  disc size and luminosity versus \mbh\ relations.

\section{Half-light radius of standard discs}
\label{sec:standard}

The disc half-light radius at wavelength $\lambda$, $R_{\rm 1/2,\lambda}$, is defined as:

\begin{equation}
    \frac{ \int_{R_{\rm ISCO}}^{R_{1/2,\lambda}} 2\pi R F_{\lambda}(R) dR}{\int_{R_{\rm ISCO}}^{\infty} 2\pi R F_{\lambda}(R) dR} =\frac{1}{2},
    \label{eq:definition}
\end{equation}
\noindent where $R_{\rm ISCO}$ is the radius of the innermost stable circular orbit (ISCO), and $F_{\lambda}(R)$ is the monochromatic flux, at wavelength $\lambda$, emitted from radius $R$. In principle, this should be the flux as detected by the observer (including all relativistic effects in the propagation of light from the disc to the observer), which is not necessarily the same as the flux emitted by the disc in its rest frame. Below (and in the next section), we discuss the half-light radius when the disc flux is cosidered under various assumptions; namely, the flux emitted in the rest frame of the disc, with and without assuming colour-correction terms, and the flux observed by a distant observer.

First, we used \kynsed\ to compute the disc flux as a function of radius and then the model half-right radius, \hlrmod, when the disc is not illuminated by the X-ray source\footnote{All half-radius model calculations in this paper are done for $\lambda=2500$~\AA, so that we compare the model with the data listed in Table \ref{table:data}. Hence, \hlrmod\ refer to the model half-radius measured at 2500~\AA.}. When X-ray illumination is turned off, \kynsed\ calculates the SED of a plane parallel, Keplerian, geometrically-thin and optically-thick accretion disc around a BH of mass \mbh\, and spin \spin$^($\footnote{\spin$=Jc/G$\mbh$^2$, where $J$ is the angular momentum of the BH. The spin parameter is smaller or equal to 1 \cite[see e.g.][]{Misner73}}$^)$, with an accretion rate of \mdot. Figure\,\ref{fig:radprofl} shows the flux radial profile in the case of an AGN with \mbh=$10^8$\msun, \mdot=0.2, \spin=0 and 1 (left and right panels, respectively). The inclination angle, $\theta$, is set to 40\degr. The red lines show the flux radial profile when the disc temperature is estimated ignoring the relativistic factors
(which is usually the case). The dotted blue lines show the flux profile when the temperature is properly computed, following the NT prescription. The radial profile is systematically smaller in the latter case, although the difference is not significant, except at radii smaller than $\sim 20$~\rg\, when \spin=0 (left panel in Fig.~\ref{fig:radprofl}). 

The solid blue line in both panels shows the flux radial profile when we take into account the colour-correction factor, $f_{\rm col}$, of \cite{Done2012}. For the BH mass and accretion rate we assumed, the NT disc temperature, \nttemp, is larger than $3\times 10^4~\rm K$ in the inner part. Therefore, $f_{\rm col}>1$ and the actual disc temperature is higher than $T_{\rm NT}(R)$ \citep[see][]{Done2012}. As a result, the disc flux at 2500~\AA\ is decreased in the inner disc, as most of the disc photons are now emitted at higher frequencies. Then, $f_{\rm col}$ arrives at unity at distances $\sim 70$ and 40 \rg, for spin 0 and 1, respectively. It is for this reason that the dotted and solid blue lines in Fig.~\ref{fig:radprofl} are identical at larger radii.  

The red and blue lines in Fig.~\ref{fig:radprofl} show the flux radial profile in the rest frame of the disc. However, a distant observer will detect the black lines (in the same panels) that show the disc flux radial profile at infinity, namely, taking into account all the relativistic effects (gravitational and Doppler shifts, aberration, light bending etc). Relativistic effects are not important in the case of a Schwarzschild BH (the black and blue lines are almost identical at all radii), but they become significant in the inner part of a maximally rotating BH (at $r \leq 6$~\rg).

The red-dashed vertical lines in Fig.~\ref{fig:radprofl} indicate the half-light radius that corresponds to the red lines. This is the half-light radius that is usually computed, without considering relativistic effects and the colour-correction factor. It is equal to 2.44$\times R_{\lambda_{\rm rest}}$, where $R_{\lambda_{\rm rest}}$ is the scale length of the disc flux profile  \citep[see e.g. Eq 2 in][]{Morgan10}. The black vertical lines in the same panels indicate \hlrmod\ in the case of a NT disc as observed at infinity (dot-dashed lines) and when we also consider a colour-correction factor (black dashed lines). 
For this particular choice of model parameters, \hlrmod\ increases by a factor of $\sim 1.35$ and 1.5, for \spin=0 and 1, respectively, when compared to \hlrmod\ that is computed analytically. 

\section{Half-light radius of X-ray illuminated discs}
\label{sec:illuminated}

We assumed an X-ray source that is point-like (lamp-post geometry) and located at a height, $h,$ above the BH, on its rotational axis. We further assume that the X-ray source emits isotropically in its rest-frame. Part of the X-rays are emitted towards the disc, where they are either absorbed or reflected. With \kynsed, we\, can calculate the SED when the X-ray luminosity is a free model parameter and when the power that heats the corona is linked to the power that is released by the accretion process. In the latter case, the power dissipated by the accretion flow below a 'transition' radius, \rtr, is assumed to be transferred to the X-ray corona. The X-ray luminosity is not a free parameter any more. Instead, it depends on \rtr\, and \mdot\, (for a given BH mass). We chose the second option to study the size of X-ray illuminated discs.

The X-ray energy spectrum is considered to be a power law with a photon index $\Gamma$, which extends from a low-energy, $E_0$, up to a high-energy cut-off, $E_{\rm cut}$. The high energy cut-off is determined by the corona temperature, and is a free parameter of the model. The low-energy cut-off is set by the mean energy of the disc photons which enter the corona. With \kynsed, we\ estimate the accretion disc spectrum taking into account the X-ray flux that is absorbed by each disc element. However, at the same time, the X-ray luminosity, hence the X-rays absorbed by the disc, depend on the spectrum of the disc photons entering the corona. \kynsed\ utilizes an iterative process to solve this impasse, where the disc spectrum and $E_0$ are computed repeatedly until $E_0$ does not change by more than 1\%\ (see D22 for details). Once $E_0$ is set, the model determines the incident X-ray flux in each disc element and it computes the ionisation state of the disc (all models in this work were computed assuming a disc density of $10^{15}$ cm$^{-3}$). This determines the amount of X-rays that will be reflected and, finally, the amount of X-rays that are absorbed by the disc (see K21 for a detailed description of this process). 

The absorbed X-ray flux is an additional source of heating for the disc. In fact, X-ray absorption is the only source of heating for the disc below \rtr\, if we assume that all the accretion power that is dissipated at these radii is transferred to the corona.  At the same time, the disc emission at radii larger than \rtr\, is enhanced because the local temperature is now higher than \nttemp\ as a result of the extra heating provided by the X-rays. 

As an example, the magenta lines in Fig.\,\ref{fig:radprof-rtrans} show the disc flux radial profile (at infinity) when the disc is illuminated by an X-ray corona at $h=20$~\rg\, (\spin=1, and the other physical parameters are the same as in Fig.\,\ref{fig:radprofl}). The transition radius is \rtr=3\risco, which implies that 0.47 of the total accretion luminosity is transferred to the corona. The black lines in this figure show the radial profile when \rtr=\risco, that is,\, the disc flux radial profile when it is not illuminated by the X-rays (as observed at infinity). 
These are the same as the black lines in the right panel of Fig.~\ref{fig:radprofl}. 

The black and magenta dotted lines in Fig.\,\ref{fig:radprof-rtrans} clearly show that the NT disc flux at all radii greater than \rtr\ increases when X-rays illuminate the disc, because the disc is hotter due to X-ray absorption. The black and magenta solid lines in the same figure show that the disc flux increases at radii larger than the radius where $f_{\rm col}=1$ ($\sim 40$~\rg, in this case). At smaller radii, down to \rtr, $f_{\rm col}>1$, and the wavelength for some of the disc photons that were originally emitted at 2500~\AA\, is now shorter. Hence, the magenta and black solid lines in Fig.\,\ref{fig:radprof-rtrans} are comparable at these radii. A discontinuity appears at $R=$\rtr\ in the case of the X-ray illuminated disc radial profiles (magenta lines in Fig.\,\ref{fig:radprof-rtrans}). This is because the disc at lower radii is heated by X-ray absorption only, and the temperature is smaller than \nttemp\ in this case.

The vertical dashed lines in Fig.~\ref{fig:radprof-rtrans} show the location of \hlrmod. The half-light radius increases when X-rays irradiate the disc, mainly because of the flux increase at large radii. The half-light radius when we also consider the colour-correction factor is almost two times larger than the half-light radius that is computed analytically (i.e. comparing the  magenta vertical dashed line in Fig.\,\ref{fig:radprof-rtrans} and the vertical red dashed line in the right panel of Fig.\,\ref{fig:radprofl}). Actually, as we show below, the half-light radius can be even larger than this factor, even for Schwarzschild BHs, when X-rays illuminate the disc. 

\section{Model disc half-light radius and luminosity}
\label{sec:model}

The main parameters of \kynsed\ are \spin, \mdot, \rtr/\risco, $\theta$, and the height of the X-ray corona, $h$. We used \kynsed\ and computed \hlrmod\ at 2500~\AA, as well as the disc luminosity at 3000~\AA\ (\lummod)\footnote{All model luminosity calculations are done for $\lambda=3000$ \AA, so that we compare the model with the flux data listed in Table \ref{table:data}, at the rest frame of each source.} in three different cases: (a) NT disc, (b) NT disc illuminated by X-rays, and (c) NT disc which is illuminated by the X-rays. We assumed the colour-correction term given by \cite{Done2012}, both before and after X-ray illumination (model NT, NT/X,  and NT/X+fcol, respectively.) In all cases, we computed the half-light radius and the disc luminosity for all combinations of the model parameter values listed in Table \ref{table:param} (there are 5250 pairs of \hlrmod\ and \lummod, for each inclination). The range of BH mass values we consider is similar to the range of the BH mass estimates for the quasars in the sample. We did not consider accretion rates smaller than a few percent or close to the Eddington accretion rate, because the NT solution may not be appropriate in these cases. We did not consider a coronal height smaller than 5\,\rg and higher than 100\,\rg, because the X-ray illumination effects are not significant in either case (see e.g.\, K21). As for \rtroverisco, we did not consider values greater than 12 because even at \rtroverisco=12 more than 70--80\%\ of the total accretion power that is dissipated to the disc is transferred to the corona, depending on spin (see Figure 1 in D22). The calculations were done assuming an X-ray spectral photon index of $\Gamma=2$ and a high energy cut-off of 300 keV (we verified that the results do not change if we assume different values of these parameters, which mainly define the shape of the X-ray spectrum). Tables \ref{table:modeldata1} and \ref{table:modeldata2} list the NT, NT+fcol, NT/X, and NT/X+fcol model half-light radius and luminosity, respectively, for all model parameters that we considered in this work.

\begin{figure}
\centering
\includegraphics[width=1\linewidth]{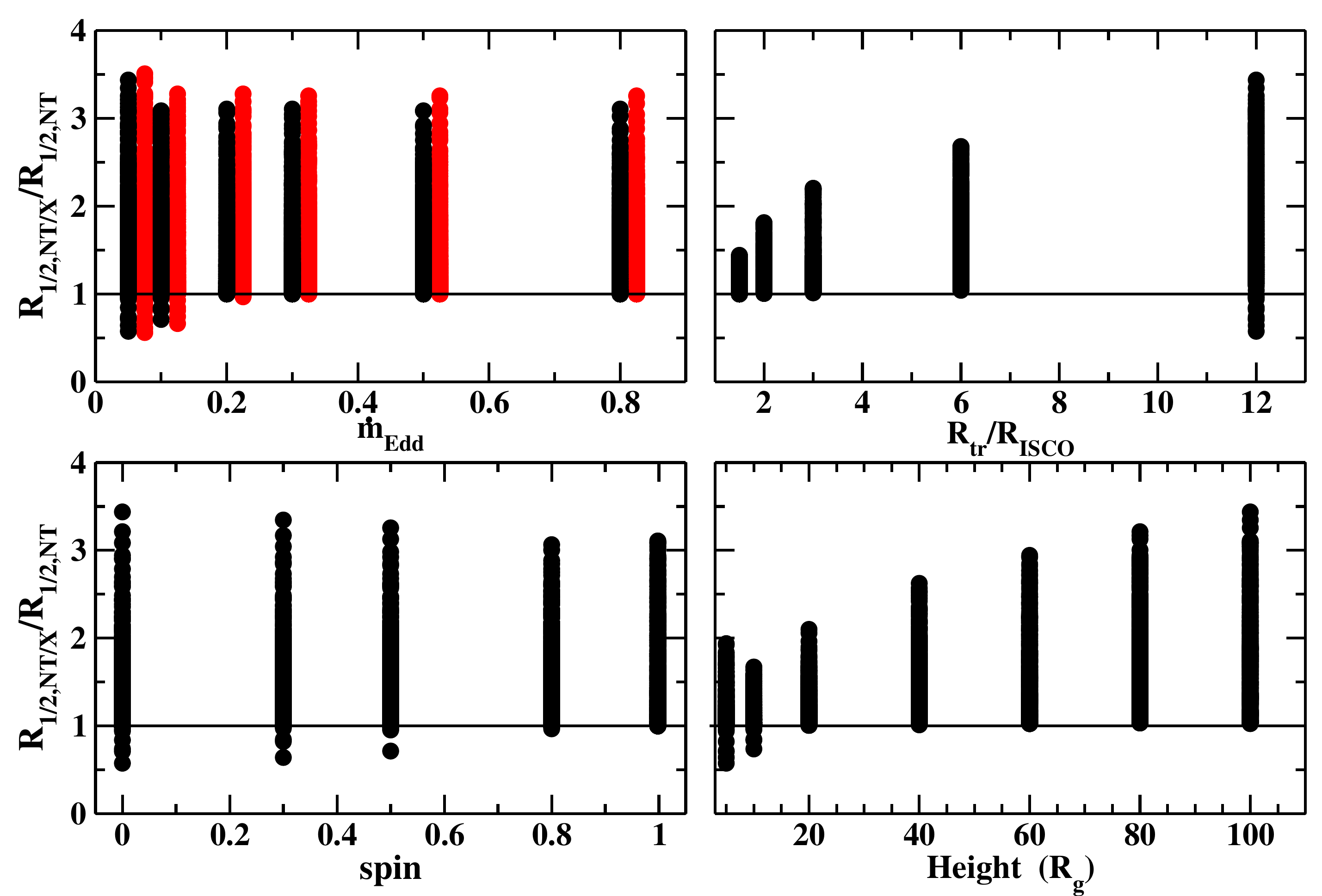}
\caption{Ratio of the half-light radius in the case when X-rays illuminate a NT disc (NT/X model) over the half-light radius in the case of the non-illuminated NT disc.}
\label{fig:comparison}
\end{figure}

Figure \ref{fig:comparison} shows the ratio of $R_{1/2,NT/X}/R_{1/2,NT}$, namely, the ratio of the half-light radius of a NT, X-ray irradiated disc over the half-light radius of a non-illuminated NT disc (for the same \mbh, \spin, \mdot, and $\theta$). The half-light radius increases by a factor up to $\sim 3.5$ when a NT disc is illuminated by the X-rays. We stress that we take into account all the relativistic effects when we compute the ratio.
This ratio would be larger if we would compare $R_{1/2}$ of an X-ray illuminated disc and the half-light radius computed with the simplified equations that are often used in practice, especially in the case of high BH spins. 

The ratio is maximized at a high \rtr\ (when the power transferred to the corona is great) and coronal height, since the fraction of the thermally reprocessed component over the total disc flux is also at maximum in this case (see Section 4.3 in K21). We note that NT X-ray illuminated discs may also appear smaller than NT discs when \rtr\ is very large and the coronal height is smaller than $\sim 10-20$~\rg. In this case, a large part of the disc is heated only from X-ray absorption. 
Since the coronal height is low, X-rays illuminate significantly the inner disc only. As a result, the flux is concentrated towards small radii and the half-light radius is smaller than the NT disc. Black and red points in the top left panel Fig.\,\ref{fig:comparison} show the ratio of $R_{1/2,NT/X}/R_{1/2,NT}$ when $\theta=30\degr$ and 60\degr, respectively. The ratio is practically the same, irrespective of the disc inclination.

\begin{table}
\centering
    \caption{Model parameter values we used to compute \hlrmod, and \lummod.}
\begin{tabular}{ll}
\hline \hline
Parameter       &               Parameter values        \\ \hline
\mbh ($\times 10^8$M$_{\odot}$)   &  0.316, 1, 3.16, 10, 31.6 \\
\spin\, &       0, 0.3, 0.5, 0.8, 0.998 \\
\mdot\, &       0.05, 0.1, 0.2, 0.3, 0.5, 0.8\\  
$h$ (in \rg)    &       5, 10, 20, 40, 60, 80, 100 \\
\rtr/\risco     &       1.5, 2, 3, 6, 12 \\
$\theta$ & 30\degr,60\degr \\ \hline
\end{tabular}
\label{table:param}
\end{table}

\begin{table*}
\centering
\caption{NT, NT+fcol,  NT/X, and NT/X+fcol model half-light radius (at 2500\AA), for all the model parameter combinations we considered. The full table is available online at the CDS.}
\begin{tabular}{clllcccccc}   
\hline \hline
\mbh &  $\theta^{\rm o}$ & \spin & $\dot{m_{\rm Edd}}$ & \rtr/\risco & $h$ & $R_{1/2}^{\rm NT}$ & $R_{1/2}^{\rm NT+fcol}$ & $R_{1/2}^{\rm NT/X}$ & $R_{1/2}^{\rm NT/X+fcol}$ \\
(10$^{8}$M$_{\odot}$) &  & & &  & (\rg)  & ($10^{14}$ cm) & ($10^{14}$ cm) & ($10^{14}$ cm) & ($10^{14}$ cm) \\ 
\hline
0.316 & 30 & 0.0 &  0.05 &      -- & -- &  3.95 & 4.87 & -- & -- \\
0.316 & 30 & 0.0 &  0.05 &      1.50 & 5 &  --  & -- & 3.98 & 4.87  \\
0.316 & 30 & 0.0 &  0.05 &  2.0  & 5 &  --      & -- & 4.03 & 4.87  \\
0.316 & 30 & 0.0 &  0.05 &  3.0  & 5 &  --      & -- & 4.14 & 4.87  \\
\end{tabular}
\label{table:modeldata1}
\end{table*}

\begin{table*}
\centering
\caption{NT, NT+fcol, NT/X, and the NT/X+fcol model luminosity (at 3000\AA). The full table is available online at the CDS.}
\begin{tabular}{clllcccccc}   
\hline \hline
\mbh &  $\theta^{\rm o}$ & \spin & $\dot{m}_{\rm Edd}$ & \rtr/\risco & $h$ & log($L^{\rm NT}_{3000\AA}$) & log($L^{\rm NT+fcol}_{3000\AA}$) & log($L^{\rm NT/X}_{3000\AA}$) & log($L^{\rm NT/X+fcol}_{3000\AA}$)\\
(10$^{8}$M$_{\odot}$) & & & & & (\rg) &   & (erg/s) & (erg/s)  & (erg/s) \\ 
\hline
0.316 & 30 & 0.0 &  0.05 &      -- & -- & 43.88 & 43.81 & --  &   -- \\
0.316 & 30 & 0.0 &  0.05 &      1.50 & 5 & -- & -- & 43.88  &   43.81 \\
0.316 & 30 & 0.0 &  0.05 &  2.0  & 5 & -- & -- & 43.88  &   43.82 \\
0.316 & 30 & 0.0 &  0.05 &  3.0  & 5 & -- & -- & 43.88  &   43.82\\
\end{tabular}
\label{table:modeldata2}
\end{table*}

\begin{figure}
\centering
\includegraphics[width=1\linewidth]{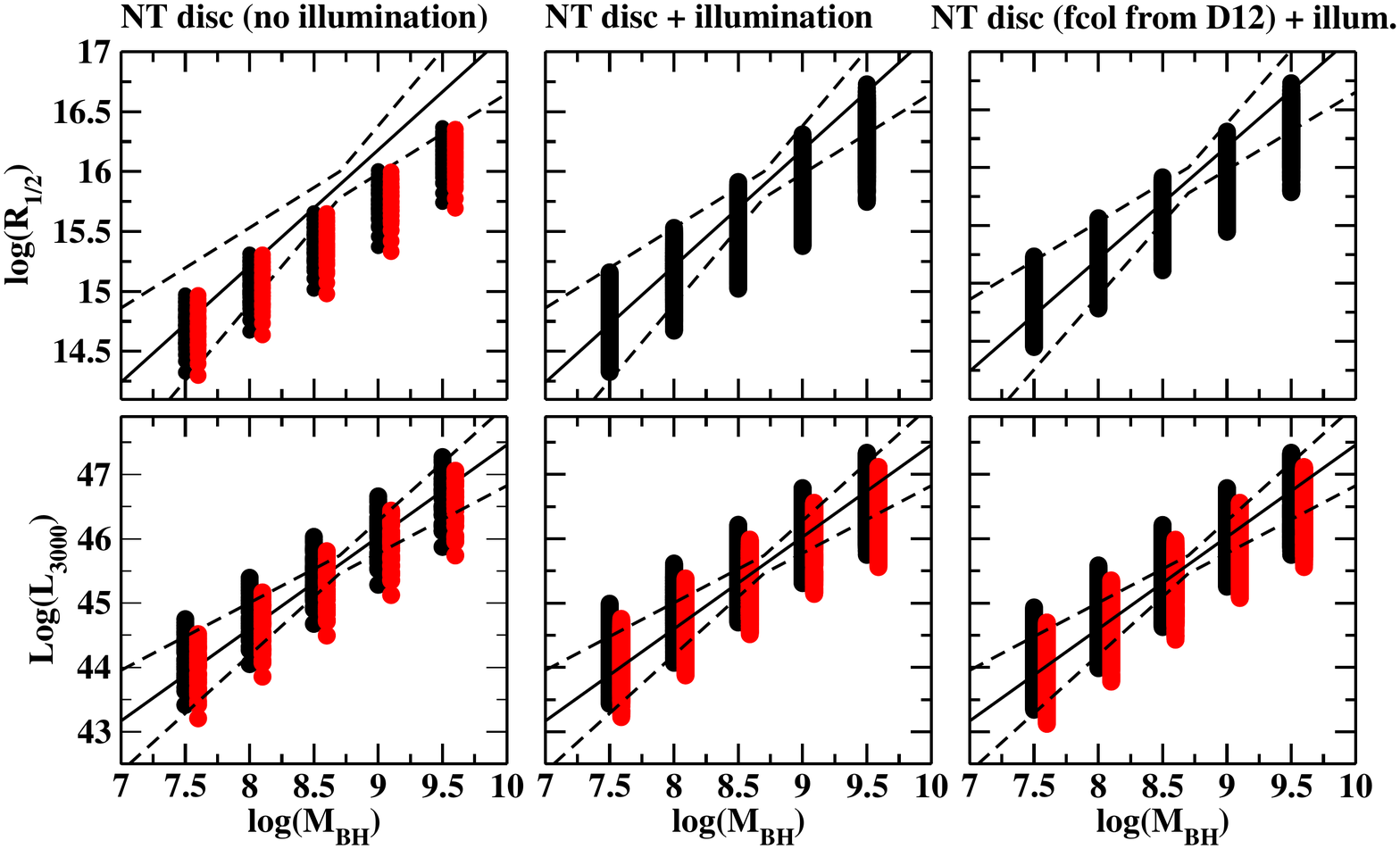}
\caption{ Plot of \hlrmod\ and of \lummod\ versus log(\mbh) (upper and lower panels, respectively) in the case of a NT disc (left panels), the NT/X model (middle panels), and the NT/X+fcol model (right panels). Black and red points in the top left panel show the half-light radius for disc inclination 30 and 60 degrees, respectively. Black and red points in the bottom panels indicate the model predictions for $\theta=30^{\circ}$ and $\theta=60^{\circ}$, respectively (\hlrmod\ are almost identical for both inclinations). Lines show the best fits to the data, and the 1$-\sigma$ confidence regions (see text for details).}
\label{fig:model}
\end{figure}

Figure~\ref{fig:model} shows plots of logarithm of \hlrmod\ (in cm) and  \lummod\ (in erg/s) as a function of log(\mbh) for the three models and all the model parameters we considered (i.e. all BH spins, X-ray luminosity, coronal height, etc). There is a positive correlation between \hlrmod\ and \lummod\ with \mbh, similar to what is observed, in all three cases. The black and red points in the top-left panel show the half-light radius in the case when the disc inclination is 30 and 60 degrees, respectively. The red points are slightly shifted to the right for clarity reasons. The half-light radius  does not depend (significantly) on the inclination. The flux profile changes, but mainly in normalization, and in such a way that the half-light radius remains constant. 
Noticeable differences appear at very large inclinations (above 80\degr) and high spins, when the flux from the inner part of the accretion disc behind the black hole is highly amplified due to strong light bending effect. Black and red points in the lower panel show \lummod\ when the disc inclination is 30\degr\ and 60\degr, respectively (red points are slightly shifted for clarity reasons). 
The black solid line in the upper panels shows the best-fit lines to the \hlrobsa\, vs \mbh\ relation. The line in the bottom panels is the (\alumotwo, \blumotwo) best-fit line (see \S\,\ref{sec:data}). 

The bottom panels in Figure \ref{fig:model} show that all three models are consistent with the observed luminosity-BH mass relation for the gravitationally lensed quasars in our sample. On the other hand, the half-light radius versus BH mass model predictions  
are flatter than the observed relation in the case of the NT model (although they are consistent within 1$\sigma$; top-left panel). The NT model predicts smaller half-light radii for sources with BH mass larger than $\sim 4\times 10^8$M$_{\odot}$. This is mainly because the disc temperature decreases with increasing BH mass (for the same accretion rate). The NT/X and NT/X+fcol models are consistent with the observations (middle and right panels, respectively). The X-ray illumination makes the disc hotter at larger radii and the half-light radius increases accordingly (see Fig.\,\ref{fig:comparison}). As a result, many model parameter combinations can result in \hlrmod--\mbh\ and \lummod--\mbh\ relations which are consistent with the observations.

\section{Comparison between models and the observations}

\subsection{Constraints from the observed half-light and luminosity versus M$_{\rm BH}$ relations}
\label{sec:results}

Each combination of \mdot, $h$, \spin, and \rtroverisco, (\mdot\, and \spin, for the NT model) results in a set of five (\hlrmod, \mbh) and (\lummod, \mbh) values, for the five BH masses that we consider here. In order to investigate whether the models are consistent with the data we need to investigate whether there is a plausible range of model parameters that can result in (\hlrmod, \mbh) and (\lummod, \mbh) sets of points, both of which are expected to be within the respective $1\sigma$ confidence regions defined by the best-fit lines to the data (as indicated by the dashed lines in Fig.~\ref{fig:model}). 

First, we identified all the parameter combinations which result in (\hlrmod, \mbh) points which are within the 1$\sigma$ confidence region for all the five BH masses. For example, in the case of the NT/X model, we identified all the (\mdot, $h$, \spin, \rtroverisco) combinations that result in \hlrmod\, values, which are within the dashed lines plotted in the top middle panel of Fig.~\ref{fig:model}, for all the five BH masses. Then, we considered the predicted \lummod\, values for these parameters and we checked whether all the respective five (\lummod\, \mbh) points are also within the 1$\sigma$ confidence region (indicated by the dashed lines in the bottom-middle  panel of Fig.~\ref{fig:model}). We accept that a model is consistent with the data at the $1\sigma$ level if there are model parameters for which the five sets of (\lummod, \mbh) and (\hlrmod, \mbh) pairs are both within the 1$\sigma$ confidence region of the best-fit lines to the data. The range of these model parameters should be representative of the 1$\sigma$ confidence region of the 'best-fit' model parameters. Our results are listed in Table \ref{table:modelparam}. 

\begin{table}
\centering
    \caption{ Parameter range for the NT,NT/X and NT/X+fcol models that result in (\hlrmod,\mbh) and (\lummod,\mbh) points which are consistent with both the observed \hlrobs\, vs \mbh\, and the \lumobs\, vs \mbh\, relations, within $1\sigma$, and $2.5\sigma$ for the NT model (parameter values listed in parenthesis for this model).}
\begin{tabular}{llllll}
\hline \hline
Model     &     $\langle \theta \rangle$ & \mdot\  &  \spin\   &   $h$(\rg)     &  \rtroverisco\    \\ \hline
NT        &  30$^{\circ}$ & -- & -- & -- & -- \\   
(NT/$2.5\sigma$)          &  (30$^{\circ}$) & (0.1-0.5) & ($\le 0.8$) & -- & -- \\
NT        &  60$^{\circ}$ & -- & -- & -- & -- \\
(NT/$2.5\sigma$)         &  (60$^{\circ}$) & (0.2-0.5) & ($\le 0.8$)  & -- & -- \\
NT/X      & 30$^{\circ}$ & $\le 0.1$ & $0-0.998$ & $\ge 40$ & $\ge 6$ \\
NT/X      & 60$^{\circ}$ & $\le 0.3$ & $\leq 0.5$ & $\ge 20$ & $\ge 6$\\
NT/X+$\rm f_{col}$ &  30$^{\circ}$ & $\le 0.2$ & $0-0.998$ & $\ge 40$ & $\ge 6$ \\
NT/X+$\rm f_{col}$ &  60$^{\circ}$ & $\le 0.3$ & $\leq 0.5$ & $\ge 40$ & $\ge 3$ \\ \hline

\end{tabular}
\label{table:modelparam}
\end{table}

There are no NT model parameters that can result in sets of (\hlrmod, \mbh) and (\lummod, \mbh) points that will simultaneously be consistent with the half-light radius and luminosity versus BH mass observations at the 1$\sigma$ level. We found that the NT model is consistent (at 1$\sigma$) with the observed half-light radius vs BH mass relation  when \mdot$\ge 0.5$ and \spin$\leq 0.3$. High accretion rates are necessary for the NT discs to explain the measured half-light radius of the quasars in the sample. However, such high accretion rate results in too high luminosity. In fact, the NT model can explain the observed luminosity of the sources in the sample (at the $1\sigma$ level), only if \mdot\ is smaller than $0.5$. In other words, the NT disc models that can explain the observed half-light radius values are 'too' bright when compared with the observed luminosity of the sources. We found that the NT model is consistent with the observations at the $2.5\sigma$ level, for the parameter values that are listed in parentheses in Table \ref{table:modelparam}.

On the other hand, X-ray illuminated discs are consistent (at 1$\sigma$) with the data. The extra power that heats the disc due to X-ray illumination can explain the observed \hlrobs\, values with lower accretion rates, thus explaining the observed luminosity at the same time. The range of the parameter values that are consistent with the data is broadly similar for both models. We cannot put strong constraints on the spin of the quasars, except in the case of NT/X and NT/X+f$_{\rm col}$ models, which predict spins smaller than 0.5. On the other hand, the illuminated disc models predict relatively low accretion rates ($\leq 0.3-0.2$). Actually, the accretion rates that we find are consistent with what is observed for sources of similar mass and redshift range \citep[see e.g.][]{Kollmeier2006, Lusso2012, Nobuta2012, Capellupo2015}. In addition, the models predict great coronal heights, which is expected, since the disc heating is more efficient when the coronal height is great (see K21). They also require that \rtroverisco $\ge 6$, which implies that at least 50\% of the accretion power that is released on the disc is transferred to the corona. We note that even in this case, the disc is quite more luminous than the X-ray corona, as half of the X-ray luminosity is directed away from the observer, towards the disc (if the X-ray source is emitting isotropically in its rest frame). 

\subsection{Constraints from the X-ray luminosity}
As a final test for the X--tray illumination models (i.e.\, NT/X-fcol and NT/X models), we computed the model X-ray luminosity and we compared it with the observations. We considered four BH masses (1, 5, 10, and $50\times 10^8~\rm M_{\odot}$), with \spin~$=0, 0.3,$ and 0.5, and three accretion rates (\mdot~$=0.01, 0.05,$ and 0.1). We considered two transition radii (\rtr~$=6$ and 12~\risco), and a coronal height of 60, 80, and 100~\rg. As we show above, these are the physical parameters that would result in half-radius--\mbh\ and luminosity--\mbh\ relations consistent with the data, in the case of X-ray illuminated discs. 

\begin{figure}
\centering
\includegraphics[width=0.95\linewidth]{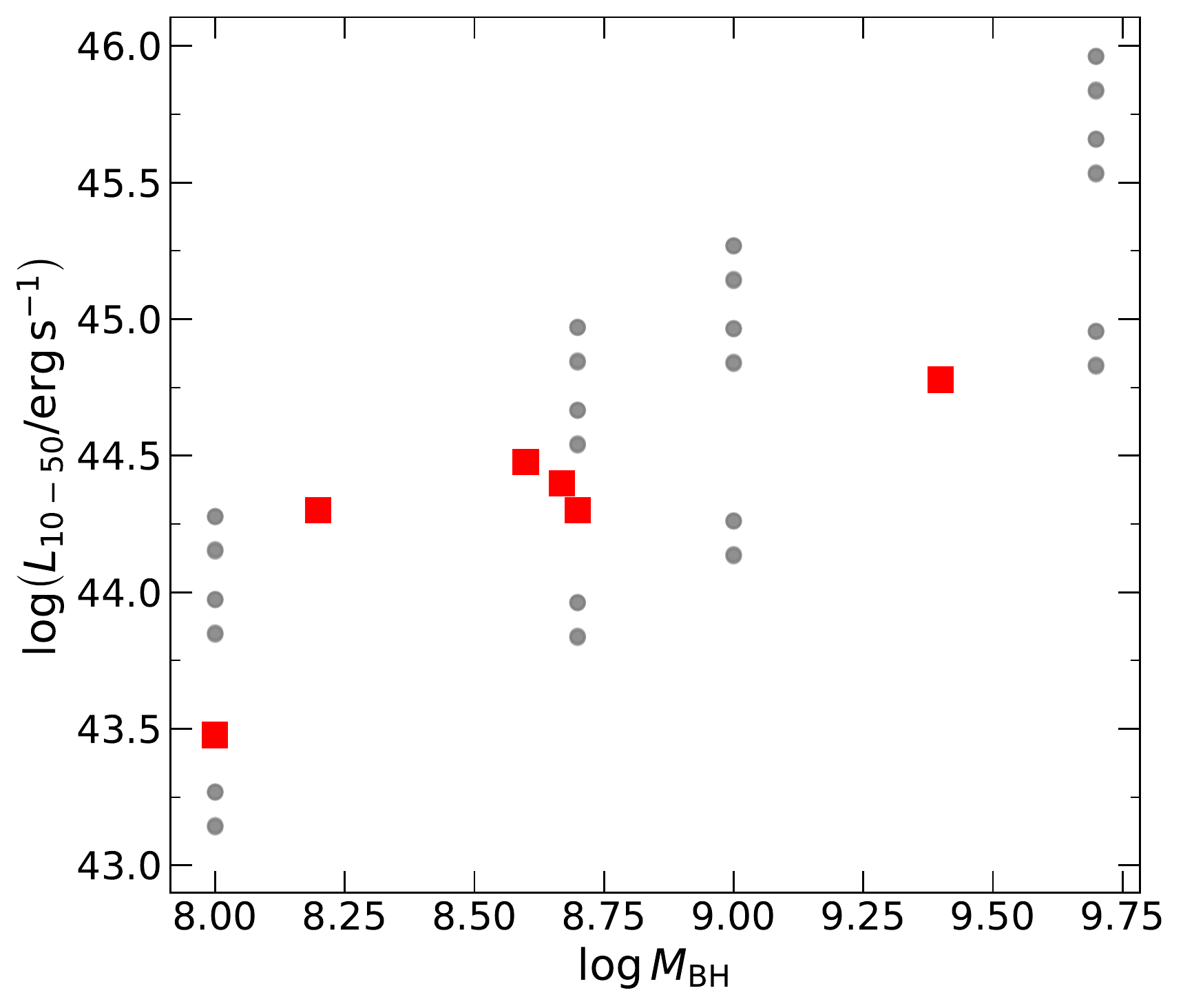}
\caption{ $10-50~\rm keV$ luminosity as a function of BH mass. The grey circles show the model predictions, estimated using {\tt KYNSED} for various combinations of height, \spin, \mdot, and \rtr. The red squares show the observed values  for QJ~0158, HE~0435, SDSS~0924, SDSS~1004, HE~1104, and Q~2237, from \cite{Chen12} (see text for details).}
\label{fig:plotlx}
\end{figure}

We ran \kynsed\ for each combination of the parameter space, and we computed the X-ray luminosity in the 10--50 keV band, \lxm\ (in ergs/s), assuming $\theta=30\degr$, $\Gamma=1.8$, and $E_{\rm cut}=150$~keV. Grey circles in Fig.\,\ref{fig:plotlx} show the results. We note that the hard X-ray luminosity is almost identical in the case of the NT/X and the NT/X+fcol models. The colour-correction factor affects the optical/UV part of the spectrum (and not the X-rays), except in cases of very low coronal height, and high spins (even in this case, the difference is less than 10\%). In general,  \lxm\ increases with BH mass, as expected. It also increases with \mdot. The \lxm\ values appear in pairs for each \mdot; the larger luminosity corresponds to the larger \rtr. The coronal height does not affect significantly the X-ray luminosity since we have considered heights larger than 10\rg\ (see Figure~A.5 in D22). The same is true for \spin\ (see Figure~A.1 in D22). 

The red points in the same figure indicate the observed X-ray luminosity in the (rest-frame) 10--50~keV band, \lxobs,  for six quasars. The measurements are taken from \cite{Chen12}, who presented the results from the analysis of {\it Chandra} monitoring data of six gravitationally lensed quasars, which also belong in our sample (namely: QJ~0158, HE~0435, SDSS~0924, SDSS~1004, HE~1104, and Q~2237). These authors measured the X-ray luminosity  correcting for magnification) in various bands. We chose to compare the model with the measurements in the 10--50~keV band, in order to minimize any complications due to the (possible) presence of absorption, either in the host or the lensing galaxy.

Figure \ref{fig:plotlx} shows that the model predictions are fully consistent with observations. We stress that the model X-ray luminosity plotted in this figure is the one that is needed to heat the disc so that the disc size and luminosity remains consistent with the observations. In other words, Fig. \ref{fig:plotlx} confirms that the sources in the sample have the necessary X-ray luminosity to heat the disc by the amount that is required to explain the observed half-light radii (and UV luminosity) of the quasars in our sample. 

\section{Discussion and conclusions}
\label{sec:discuss}

We studied the half-light radius versus BH mass and the disc luminosity versus BH mass relations in the case of a NT accretion disc as well as a NT disc that is illuminated by X-rays. We also considered the case of a colour-correction factor, as proposed by \cite{Done2012}. We used {\tt KYNSED}, which is a recently developed model for the broad-band SED of AGN. It computes the SED in the case when the X-ray corona is located on the spin axis of the BH, and illuminates the disc. It takes into account all the GR effects, as well as the disc ionisation, when computing the X-ray reflection spectra, and, hence, the amount of the X-rays that are absorbed by the disc. 

One of our main results is that the half-light radius can increase by a factor of up to $\sim 3.5$ when the disc is illuminated by X-rays, even when \spin=0. This is because a large part of the incident X-rays is absorbed and acts as an extra source of power to heat the disc. Consequently, the disc temperature is higher than the temperature predicted by the standard NT disc theory and the flux profile is shifted to larger radii. 

 Figure \ref{fig:comparison}  shows that the half-light radius of a NT disc can increase by a factor of $\sim 3$  if \rtr/\risco$\ge 6$. Such a transition radius implies that $\sim 55-65\%$ of the power that is released by accretion in the disc is transferred to the corona (for spin zero and one, respectively). We note that even if most of the power is transferred to the corona, the observed disc luminosity is expected to be larger than the X-ray luminosity -- if the corona emits isotropically in its rest frame. In this case, at least half of the corona luminosity is emitted towards the disc, hence, it is absorbed and heats the disc, thereby increasing its luminosity. 

A significant increase in the value of $R_{1/2}$ is not expected every time \rtr/\risco$\ge 6$, as it depends on the other physical parameters of the system. For example, the bottom-right panel in the same figure shows that for a significant increase in the half-light radius of the order of 2.5 or more, it is necessary for  \rtr/\risco$\ge 6$ {\it } and for the coronal height to be greater than 40\rg.  As the coronal height increases, the amount of X-ray illumination (and, hence, the X-ray absorption as well) increases. This is due to projection effects, which compensate any losses due to the increased distance between the corona and the outer disc (see K21 for a more detailed discussion on this effect). Consequently, the amplitude of the disc temperature profile (and, hence, $R_{1/2}$) increases with increasing height. 

We conducted a detailed study of the model $R_{1/2}-$\mbh\ and $L_{3000}-$\mbh\ plots for a wide range of parameters (listed in Table \ref{table:param}) and we compared them with the observed `half-light radius versus BH mass' and `luminosity versus BH mass' relations, using data for 15 quasars with available microlensing variability observations. We found that non-illuminated NT accretion disc models are in agreement with the observations but only at the 2.5$\sigma$ level. We get a better agreement with the observations if we assume that the disc is illuminated by the X-rays. In this case, NT discs are consistent with the observations, within $1\sigma$, irrespective of whether we assume a colour-correction factor (or not) and of whether the discs are preferentially seen face-on or whether the sources are randomly distributed at all inclinations. Our best-fit results are listed in Table \ref{table:modelparam}. In both NT/X and NT/X+fcol models, the results indicate that the accretion rate of the sources is less than $\sim 0.3$ of the Eddington limit, \rtr/\risco$\ge 6$, and the coronal height is larger than $\sim 40$\rg. This lower limit on the coronal height is probably model dependent. The disc in the current model is parallel plane. The ratio of the disc height over disc radius is small in the standard disc models (i.e. it is not zero). Therefore, as the radius increases, the disc height is also expected to increase, leading to a `flaring' disc in the outer parts. We expect smaller coronal heights to illuminate the disc more efficiently in this situation and we plan to investigate this issue in the future.

Regarding the colour-correction factor, we considered the prescription of \cite{Done2012}. The $f_{\rm col}$ factor is related to the disc opacity, which is determined by the structure of the NT disc. Therefore, it is not clear what the correct colour correction factor is when the disc is illuminated by X-rays. The fact that both a NT disc and a disc with the given $f_{\rm col}$ prescription is consistent with the observations when illuminated by X-rays indicates that the exact colour-correction factor does not affect the main conclusion of our results -- namely, that X-ray illumination of the accretion disc in AGN is needed to explain the observations.

The current sample of quasars with measured half-light radius is small. Given the size of the sample, we cannot make a general conclusion about the importance of X-ray illumination in AGN. This should depend on the amount of the power that is transferred to the corona as well as on the coronal height. These parameters must be related to the mechanism that creates and heats the X-ray corona. It is possible that \rtr/\risco\, and the coronal height will be the same in all quasars. However, since the mechanism is currently unknown, we cannot be certain of the above. It may also be possible that depending on the accretion rate, there may be a wide range of \rtr/\risco\, and coronal heights in AGN. This would imply that X-ray illumination may not be important in other AGN, but we cannot make solid predictions based on the current work. 

We note that we have not considered the possibility of internal absorption (in the optical/UV bands). For example, \cite{Gaskell2004} estimate that the typical reddening for a face-on AGN can be up to $E(B-V)\sim 0.3$. Although this may be an overestimate of the reddening for the typical AGN \citep[see e.g.][]{Baron2016}, we can assume this value just to give an example as to how absorption can influence our results. The upper limit of \mdot$\sim 0.2$ that we get in the case of $\theta=30^{\circ}$ for the NT/X-fcol model, is mainly set by the observed luminosity at 3000\,\AA. Since the disc luminosity is proportional to \mdot, a higher accretion rate will predict luminosity larger than observed. However, if we assume that the observed luminosity at 3000\AA\ is $\sim 2-2.5$ times lower than the intrinsic one (due to absorption), then the intrinsic accretion rate could be up to \mdot$\sim 0.4-0.5$. Such absorption would imply an extinction of $A_{3000}\sim 0.75-1$, respectively. Assuming the quasar extinction curve of \cite{Czerny04}, this would correspond to an $E(B-V)$ of $0.14-0.18$.

In summary, when the disc is illuminated by X-rays, the disc emission in the UV/optical is consistent with a multicolour blackbody emission, with a temperature profile determined by the accretion power dissipated in the disc plus the power due to X-ray absorption. As a result, the disc temperature increases at larger radii, when compared to the standard-disc model temperature, and the disc appears to be `larger' than expected. This effect can explain the apparent discrepancy between the disc size as determined by microlensing observations and the luminosity-black hole mass relation of gravitationally lensed quasars. 

The X-ray illumination of the disc can also explain the apparent discrepancy between the standard disc model and the observed UV/optical time-lags  measured from recent monitoring campaigns \citep[see][]{Kammoun21b}.  There are some differences between the model used by \cite{Kammoun21b}  and the model we use in the current work. The X-ray luminosity was a free parameter in \cite{Kammoun21b}, and was not part of the power that is available for the disc heating due to the accretion process. In addition, \cite{Kammoun21b} assumed $f_{\rm col}=2.4$, while we assume the colour-correction factor of \cite{Done2012}. We are currently developing a new code to fit the optical/UV time lags, for any combination of the X-ray heating source, spin, and colour-correction term. These results  will be published soon. Our preliminary results show that we can fit the optical/UV time-lags assuming X-ray disc illumination and the same assumptions we adopt in this work. 

Our results offer strong support to the X-ray disc illumination hypothesis in AGN. We plan additional tests of the disc X-ray illumination hypothesis by considering broadband AGN SEDs (in particular, SEDs that include short wavelength data). Although it is not straightforward to use \kynsed\ to fit SEDs for individual AGN if they have been constructed using contemporaneous data (see the discussion in \S\,5.5 in D22), we plan to study AGN composite SEDs published over the past few years, as well as the observed $L_{\rm 2 keV}$ vs $L_{\rm UV}$ relation (or, equivalently the $\alpha_{\rm ox}$ vs $L_{2500 \AA}$ relation) in quasars. These tests ought to provide further constraints on the current model and may aid in developing it further.

\begin{acknowledgements}

We thank H. Netzer, the referee, for his helpful and constructive comments that helped us improve the manuscript significantly. IEP acknowledges support by the project ``Support of the international collaboration in astronomy (ASU mobility)'' with the number: CZ.$02.2.69/0.0/0.0/18-053/0016972$. ASU mobility is co-financed by the European Union. MD thanks for the support from the GACR project 21-06825X and the institutional support from RVO:67985815. ESK acknowledges financial support from the Centre National d’Etudes Spatiales (CNES).

\end{acknowledgements}

%
%

\balance

\bibliographystyle{aa} 
\bibliography{references} 
%
%


\end{document}